\documentclass{article}
\usepackage{frascatiphys,here,graphicx,subfigure}
\usepackage{bm}
\usepackage{amsmath,amssymb}
\begin{document}
\title{ 
ON THE EXISTENCE OF LIGHT-SCALAR MESONS \\
$\bm{\kappa(800)}$ and $\bm{\kappa'(1150)}$: \\
THE $\bm{\widetilde{U}(12)}$ SCHEME AND BES II DATA
}
\author{
Kenji Yamada and Tomohito Maeda \\
{\em Department of Engineering Science, Junior College Funabashi Campus,} \\
{\em Nihon University, Funabashi 274-8501, Japan}
}
\maketitle
\baselineskip=11.6pt
\begin{abstract}
We present that there should exist a light strange-scalar meson
 $\kappa'$, in addition to the $\kappa(800)$,
 which has a mass around 1.1-1.2 GeV, a rather
 narrow width, and couples strongly to
 $\kappa(800)\sigma(600)$ ($K\pi\pi\pi$) but weakly
 to $K\pi$, based upon the $\widetilde{U}(12)$-classification
 scheme of hadrons and BES II data on
 $J/\psi \rightarrow \bar{K}^{*}(892)^{0}K^{+}\pi^{-}$ decay.

\end{abstract}
\baselineskip=14pt
%
%section1
\section{Introduction}
Recently, the existence of the light-scalar mesons, $\sigma(600)$
 and $\kappa(800)$, has been confirmed by showing the presence of
 respective poles in the $\pi\pi$\cite{CCL,P2004} and
 $K\pi$\cite{P2004,DM} scattering amplitudes,
 in addition to results of Breit-Wigner fits to $D$- and $J/\psi$-decay
 data, respectively, from the E791\cite{E791} and
 BES\cite{BES2004,BES2006} collaborations.
 However, the nature of these resonances, together with the
 $f_{0}(980)$ and $a_{0}(980)$, has been a long-standing problem in
 controversy, where it is not obvious how these light-scalar
 mesons are understood in terms of quarks and gluons in QCD.

Here we focus on the strange scalar mesons and
 discuss the existence of an extra $\kappa'$ meson, in addition to the
 normal $\kappa(800)$, and their strong decay properties.

%section2
\section{Existence of the extra $\bm{\kappa'}$ meson}
%
%section2.1
\subsection{The $\widetilde{U}(12)$-classification scheme of hadrons}
\sloppy
The $\widetilde{U}(12)$-classification scheme
 of hadrons,\cite{IIM2000,II2002} which has a manifestly
 covariant framework of
 $\widetilde{U}(12)_{SF} \times O(3,1)_{L}$,
 generalized covariantly from nonrelativistic 
 $SU(6)_{SF} \times O(3)_{L}$ by boosts, separating the spin and space
 degrees of freedom,
 gives covariant quark representations
 for composite hadrons with definite Lorentz and chiral
 transformation properties.
 The $\widetilde{U}(12)$-classification scheme
 has a ``static'' unitary $U(12)_{SF}$ spin-flavor symmetry
 in the rest frame of hadrons, embedded in the covariant
 $\widetilde{U}(12)$-representation space, where
 $\widetilde{U}(12)$ has as its subgroups
 the pseudounitary homogeneous Lorentz group for Dirac spinors
 and unitary symmetry group for light-quark flavors,
 \begin{equation}
	\widetilde{U}(12)_{SF} \supset \widetilde{U}(4)_{D} \times U(3)_{F}.
 \end{equation}
 Since
 \begin{subequations}
	\begin{align}
		U(12)_{SF} \supset U(4)_{D} \times U(3)_{F}, \\
		U(4)_{D} \supset SU(2)_{\rho} \times SU(2)_{\sigma},
	\end{align}
 \end{subequations}
 the static $U(12)_{SF}$ symmetry includes
 both the nonrelativistic  spin-flavor $SU(6)_{SF}$
 and chiral $U(3)_{L} \times U(3)_{R}$ symmetry\footnote{Hadron states
 are classified, aside from flavors, by the quantum numbers
 $\rho,S,L,J,P$, where $\rho$ is the net quark $\rho$-spin
 concerning $SU(2)_{\rho}$, $S$ the ordinary net quark $\sigma$-spin,
 $L$ the total quark orbital angular momentum, and
 $J$ and $P$ the total spin and parity of hadrons.}
 as
 \begin{subequations}
	\begin{align}
		U(12)_{SF} &\supset SU(6)_{SF} \times SU(2)_{\rho}, \\
		U(12)_{SF} &\supset U(3)_{L} \times U(3)_{R} \times SU(2)_{\sigma},
	\end{align}
\end{subequations}
 where $SU(2)_{\rho}$ and $SU(2)_{\sigma}$ are
 the Pauli-spin groups concerning the boosting and
 intrinsic spin rotation, respectively, of constituent
 quarks, being connected with decomposition of Dirac
 $\gamma$-matrices, $\gamma = \rho \times \sigma$.
 This implies that the $\widetilde{U}(12)$-classification
 scheme is able to incorporate effectively,
 according to dynamical consequences of QCD, the effects
 of chiral symmetry and its spontaneous breaking, essential
 for understanding of properties of the low-lying hadrons,
 into what is called a constituent quark model.

In the $\widetilde{U}(12)$-classification scheme there are two
 light-scalar meson multiplets, normal $S^{(N)}$ and
 extra $S^{(E)}$ with
 $J^{PC} = 0^{++}$ and $0^{+-}$, respectively,
 in the ground level ($L=0$).
 These $N$- and $E$-scalar multiplets
 are the chiral partners, respectively, of the $N$- and
 $E$-pseudoscalar multiplets and they form linear representations
 of the $U(3)_{L} \times U(3)_{R}$ chiral symmetry.
 Concerning the strange scalar mesons, now we have two
 $\kappa$ mesons, $\kappa^{(N)}(0^{++})$ and
 $\kappa^{(E)}(0^{+-})$. Note that the observed $\kappa(800)$
 and missing $\kappa'$ are generally mixtures of them.\cite{yamada}

%section2.2
\fussy
\subsection{The BES II data}
\begin{figure}%[H]
    \begin{center}
%\subfigure[]
        {\includegraphics[width=120mm]{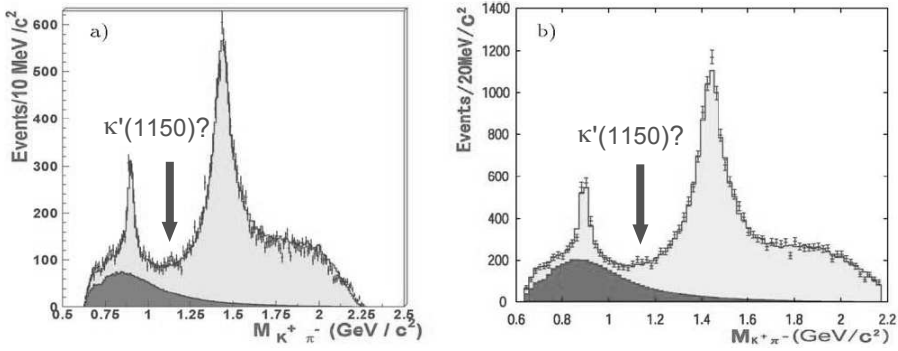}}
%\subfigure[]
        \caption{\it The $K^{+}\pi^{-}$ invariant mass spectrum in
 $J/\psi \rightarrow \bar{K}^{*}(892)^{0}K^{+}\pi^{-}$ decay
 from BES II.\cite{BES2006} The dark shaded histograms show contribution
 from the $\kappa(800)$.}
\label{BESIIkpi}
    \end{center}
\end{figure}

The $K^{+}\pi^{-}$ mass spectrum in
 $J/\psi \rightarrow \bar{K}^{*}(892)^{0}K^{+}\pi^{-}$
 decay observed by the BES II experiment\cite{BES2006}
 is shown in fig.\ref{BESIIkpi} where
 there seems to be a visible bump structure at 1.1-1.2 GeV.
 If this structure is attributed to the existence of
 a new $K\pi$ resonance, its spin-parity will likely
 be $0^{+}$ or $1^{-}$, since higher spins are unfavorable
 for such a low-mass state, and also its width is supposed
 to be narrow, judging from the data structure.
 
We hereafter refer to the strange scalar meson mentioned
 above as the $\kappa'(1150)$.\footnote{A recent lattice-QCD
 study on light-scalar mesons by the UKQCD collaboration\cite{UKQCD}
 suggests that the $a_{0}(980)$ is predominantly a conventional
 $\bar{q}q$ state, while the $\kappa(800)$ is too light to
 be assigned to the $\bar{q}q$ state, which is expected to
 have a mass about 100-130 MeV heavier than the $a_{0}(980)$.}

%section3
\section{Strong decays of the $\bm{\kappa(800)}$
 and $\bm{\kappa'(1150)}$ mesons}
We examine strong two-body decays of the $\kappa(800)$
 and $\kappa'(1150)$ as mixtures of the  $\kappa^{(N)}$
 and $\kappa^{(E)}$ in the $\widetilde{U}(12)$-classification scheme
 as follows:
\begin{equation}
	\kappa(800) \rightarrow K + \pi
\end{equation}
and
\begin{subequations}
\begin{align}
	\kappa'(1150) &\rightarrow K + \pi, \\
				&\rightarrow K + \eta, \\
				&\rightarrow \kappa(800) + \sigma(600) \ \
	[\rightarrow K\pi\pi\pi].
\end{align}
\end{subequations}
 In the actual calculations of decay matrix elements we treat
 the strange mesons $K$, $\kappa$ and $\kappa'$
 as quark-composite $n\bar{s}$ states, while $\pi$, $\eta$
 and $\sigma$ as external local fields.

%section3.1
\subsection{Quark-pseudoscalar and quark-scalar couplings}
For the effective quark-pseudoscalar coupling
 inside hadrons we assume the two independent interactions
 of the forms
\begin{subequations}
\begin{alignat}{2}
	& g_{ps}\bar{q}(-i\gamma _{5})q\phi _{p} 
	& \qquad & \text{for pseudoscalar type}, \\
	& g_{pv}\bar{q}(-i\gamma _{5}\gamma _{\mu})q\partial _{\mu}\phi _{p}
	& \qquad & \text{for pseudovector type}.
\end{alignat}
\end{subequations}
 The effective quark-scalar coupling is simply
 related to the quark-pseudoscalar coupling,
 assuming the $\sigma$ meson
 is a chiral partner of the $\pi$ meson in the linear
 representation of chiral symmetry,\footnote{Here we take chiral
 $SU(2)_{L} \times SU(2)_{R}$ as imposed symmetry.} and given as
\begin{subequations}
\begin{align}
	& g_{ps}\bar{q}q\phi _{\sigma}, \\
	& g_{pv}\bar{q}\gamma _{\mu}q\partial _{\mu}\phi _{\sigma}.
\end{align}
\end{subequations}
Then the matrix elements for the pseudoscalar($\pi,\eta$)-emitted
 processes are generally given by
\begin{subequations}
\begin{equation}
	T^{(P)} = T_{ps}^{(P)} + T_{pv}^{(P)} 
\end{equation}
 with
\begin{align}
	T_{ps}^{(P)} &= g_{ps}\big\langle \overline{W}(v')
	(-i\gamma _{5}\phi _{p})
	W(v)iv \cdot \gamma \big\rangle¡¡I_{G}^{(P)}(q^{2}) + \text{c.c.}, \\
	T_{pv}^{(P)} &= g_{pv}\big\langle \overline{W}(v')
	(-\gamma _{5}\gamma _{\mu}q_{\mu}\phi _{p})
	W(v)iv \cdot \gamma \big\rangle¡¡I_{G}^{(P)}(q^{2}) + \text{c.c.},
\end{align}
\end{subequations}
 where $W(v)$ and $\overline{W}(v')$ are the spin wave functions
 of initial- and final-state mesons,\cite{IIM2000}
 $I_{G}^{(P)}(q^{2})$ a space-time part of the Lorentz-invariant
 transition form factor,\footnote{This is obtained
 from overlap integral of space-time wave functions for
 initial and final mesons. In the present analysis we simply take
 $I_{G}^{(P)} = I_{G}^{(\sigma)} = 1$, since initial and final mesons
 both belong to the ground-state ($L=0$) multiplet in the
 $\widetilde{U}(12)_{SF} \times O(3,1)_{L}$-classification scheme.}
 $v$ and $v'$ the 4-velocities,
 $q_{\mu}$ the momentum of emitted pseudoscalar mesons, and
 $\langle \cdots \rangle$ means the trace taken over the
 spinor and flavor indices.

The matrix elements for the $\sigma$-emitted processes are
 given likewise by
\begin{subequations}
\begin{equation}
	T^{(\sigma)} = T_{ps}^{(\sigma)} + T_{pv}^{(\sigma)}
\end{equation}
 with
\begin{align}
	T_{ps}^{(\sigma)} &= g_{ps}\big\langle \overline{W}(v')(\phi _{\sigma})
	W(v)iv \cdot \gamma \big\rangle¡¡I_{G}^{(\sigma)}(q^{2}) + \text{c.c.}, \\
	T_{pv}^{(\sigma)} &= g_{pv}\big\langle \overline{W}(v')(-i\gamma _{\mu}
	q_{\mu}\phi _{\sigma})W(v)iv \cdot \gamma \big\rangle
	I_{G}^{(\sigma)}(q^{2})+ \text{c.c.}.
\end{align}
\end{subequations}

%section3.2
\subsection{Evaluation of the coupling constants}
\sloppy
The coupling constants $g_{ps}$ and $g_{pv}$ were evaluated
 by Maeda et al..\cite{maeda} They calculated  the
 $D$-wave/$S$-wave amplitude ratio and width of
 $b_{1}(1235) \rightarrow \omega(782) + \pi$ decay and obtained the values
 $g_{ps} = 2.07$ GeV and $g_{pv} = 14.0$, using
 their experimental values $D/S=0.277 \ (\pm 0.027)$ and
 $\Gamma[\omega\pi] \approx \Gamma^{\text{tot}} = 142 \ (\pm 9)$
 MeV\cite{PDG2007pu} as input.

\fussy
We now calculate the decay width of $K^{*}(892) \rightarrow
 K + \pi$ to see the validity of the present decay model
 and obtain a reasonable value of $\Gamma[K\pi] = 58$ MeV,
 compared with the experimental value $\Gamma[K\pi] \approx
 \Gamma^{\text{tot}} = 50.8 \pm 0.9$ MeV.\cite{PDG2007pu}

%section3.3
\subsection{Strong decay widths of the $\kappa(800)$
 and $\kappa'(1150)$}
Since the $\kappa(800)$ and $\kappa'(1150)$ are generally
 mixtures of $\kappa^{(N)}$ and $\kappa^{(E)}$, 
 we introduce the mixing angle $\theta$, which is
 the only free parameter in the present analysis, by
\begin{subequations}
\begin{alignat}{3}
	|\kappa(800)\rangle &= &\cos\theta \ |\kappa^{(E)}\rangle &+
	\sin\theta \ |\kappa^{(N)}\rangle, \\
	|\kappa'(1150)\rangle &= -&\sin\theta \ |\kappa^{(E)}\rangle &+
	\cos\theta \ |\kappa^{(N)}\rangle.
\label{mixing}
\end{alignat}
\end{subequations}
 Here we take the mixing angle $\theta$ to be around
 $-65^{\circ}$ so that the $\kappa(800)$ has a width of several
 hundred MeV and the $\kappa'(1150)$ a rather narrow width,
 in conformity with their observed properties, $\Gamma[\kappa(800)]
 = 550 \pm 34$ MeV\cite{PDG2007pu} and the BES II data mentioned
 above for the $\kappa'(1150)$.

Using the mixing angle $\theta = -65^{\circ}$, we evaluate
 the partial decay widths of the $\kappa(800)$ and $\kappa'(1150)$
 for respective channels in the following.

%section3.3.1
\subsubsection{Decay of the $\kappa(800)$}
If we take a mass of the $\kappa(800)$ to be 800 MeV,
 we obtain $\Gamma[K\pi] = 354$ MeV for
 $\kappa(800) \rightarrow K + \pi$. This is consistent,
 though somewhat small, with the experimental value
 $\Gamma^{\text{tot}} \approx \Gamma[K\pi] = 550 \pm 34$
 MeV.

%section3.3.2
\subsubsection{Decays of the $\kappa'(1150)$}
We take tentatively 1150 MeV for a mass of the missing
 state $\kappa'(1150)$ and then obtain
	\begin{subequations}
	\begin{alignat}{5}
	&\Gamma[K\pi]& &= 1&8& \ \text{MeV}& \ \ 
	&\text{for} \ \ \kappa'(1150) \rightarrow K + \pi, \\
	&\Gamma[K\eta]& &= &2& \ \text{MeV}& \ \ 
	&\text{for} \ \ \kappa'(1150) \rightarrow K + \eta, \\
	&\Gamma[\kappa\sigma]& &= 3&0& \ \text{MeV}& \ \ 
	&\text{for} \ \ \kappa'(1150) \rightarrow \kappa(800) + \sigma(600),
	\end{alignat}
	\end{subequations}
 where mass values of the $\kappa(800)$ and $\sigma(600)$
 are taken tentatively to be 600 MeV and 350 MeV, respectively,
 and the singlet-octet mixing angle for the pseudoscalar mesons
 $\eta$ and $\eta'$ to be $\theta _{P} = - 11.5^{\circ}$.
 From these partial decay widths we could estimate the total width
 to be
	\begin{equation}
		\Gamma^{\text{tot}} \approx
		\Gamma[K\pi] + \Gamma[K\eta] + \Gamma[\kappa\sigma]
		= 50 \ \text{MeV}.
	\end{equation}
 It may be worthwhile to mention that the dominant
 decay mode of the $\kappa'(1150)$ is not $K\pi$
 but $\kappa(800)\sigma(600)$ ($K\pi\pi\pi$), the $K\pi$
 branching ratio is $\Gamma[K\pi]/\Gamma^{\text{tot}} \approx 0.36$,
 and therefore the $\kappa'(1150)$ is supposed not to be
 seen in the $K\pi$ scattering processes.\footnote{This is consistent
 with experimental data\cite{LASS} on the $S$-wave phase of
 the $K\pi$ scattering amplitude displaying
 no typical resonance-like behavior around the energy region
 1.1-1.2 GeV.}

Here it goes without saying that the present treatment
 of the $\kappa(800)$ and $\sigma(600)$ as narrow resonances
 is quite a rough approximation and the evaluated
 decay width to $\kappa(800)\sigma(600)$ does not really make
 sense. In practice we should perform a dynamical
 calculation of the decay chain $\kappa'(1150)
 \rightarrow \kappa(800)\sigma(600) \rightarrow
 K\pi\pi\pi$, taking into account the effects of
 the broad $\kappa$ and $\sigma$ widths.

%section4
\section{Concluding Remarks}
We presented that there should exist an extra $\kappa'$
 meson which has a mass around 1.1-1.2 GeV, a rather
 narrow width, and couples strongly to
 $\kappa(800)\sigma(600)$ ($K\pi\pi\pi$) but weakly
 to $K\pi$, based upon the $\widetilde{U}(12)$-classification
 scheme and BES II data.

The strong coupling to
 $\kappa(800)\sigma(600)$ suggests that
 to observe the $\kappa'$ meson experimentally it might be favorable
 to study the $K\pi\pi\pi$ system, for example, in
 $J/\psi \rightarrow K^{*}(892)(K\pi\pi\pi)$ decay and
 $e^{+}e^{-} \rightarrow K^{*}(892)(K\pi\pi\pi)$ processes.
 However, if the main component of $\kappa'$
 is $\kappa^{(E)}(0^{+-})$, as is in the present analysis
 (taking $\theta = -65^{\circ}$ in eq.\ref{mixing}),
 the $\kappa'$ production in these processes\footnote{
 The $J/\psi$ and virtual photon from $e^{+}e^{-}$ annihilation
 have negative $C$ parity.
 The $\kappa'(1150)$ production in the
 $J/\psi \rightarrow \bar{K}^{*}(892)^{0}K^{+}\pi^{-}$ decay process
 is doubly suppressed by $C$ invariance in the $SU(3)_{f}$ limit
 and its small $K\pi$ branching ratio. This suppression
 coincides with the experimental data
 of quite small production of the $\kappa'(1150)$
 compared to the $\kappa(800)$,
 as is seen in fig.\ref{BESIIkpi}.}
 would be suppressed by charge-conjugation ($C$) invariance
 in the limit of $SU(3)_{f}$ symmetry.\cite{HP1985}
 Rather, the $\chi_{c0,1,2} \rightarrow K^{*}(892)(K\pi\pi\pi)$
 decay processes would be more promising, since they are $C$-parity
 allowed decays in the $SU(3)_{f}$ limit.

In a future study it is necessary to calculate
 dynamically the decay $\kappa' \rightarrow
 \kappa(800)\sigma(600) \rightarrow K\pi\pi\pi$
 in order to obtain a more realistic decay width
 and also to make the present decay model more effective
 by examining various strong decay processes.

%references

%
\end{document}